\documentclass[12pt]{article}
\usepackage[margin=1in]{geometry}
\usepackage[round]{natbib}
\usepackage{amsmath,amssymb,authblk,graphicx,hyperref}

\DeclareMathOperator*{\argmax}{arg\,max}

\def\bI{\mathbf{I}}

\newtheorem{proposition}{Proposition}
\newtheorem{theorem}{Theorem}
\newtheorem{assumption}{Assumption}

\title{Nonparametric false discovery rate control for identifying simultaneous signals}
\author[1]{Sihai Dave Zhao}
\affil[1]{Department of Statistics, University of Illinois at Urbana-Champaign}
\author[2]{Yet Tien Nguyen}
\affil[2]{Department of Mathematics \& Statistics, Old Dominion University}

\begin{document}
\maketitle

\begin{abstract}
  It is frequently of interest to jointly analyze multiple sequences of multiple tests in order to identify simultaneous signals, defined as features tested in multiple studies whose test statistics are non-null in each. In many problems, however, the null distributions of the test statistics may be complicated or even unknown, and there do not currently exist any procedures that can be employed in these cases. This paper proposes a new nonparametric procedure that can identify simultaneous signals across multiple studies even without knowing the null distributions of the test statistics. The method is shown to asymptotically control the false discovery rate, and in simulations had excellent power and error control. In an analysis of gene expression and histone acetylation patterns in the brains of mice exposed to a conspecific intruder, it identified genes that were both differentially expressed and next to differentially accessible chromatin. The proposed method is available in the R package \url{github.com/sdzhao/ssa}.
\end{abstract}

\section{\label{sec:intro}Introduction}
Methods for controlling the false discovery rate of a large number of hypothesis tests are now essential to many areas of scientific research. Most existing methods are intended for finding non-null signals within a single sequence of multiple tests. For example, a common application in genomics is to identify which genes, among possibly tens of thousands of candidates, are truly associated with some phenotype of interest. With the ready availability of large amounts of data, however, it has become easier to collect multiple sequences of multiple tests. For example, the same genes or genetic variants may be tested in multiple independent experiments. Each experiment then gives rise to a separate sequence of multiple tests. Jointly analyzing these multiple sequences can provide important scientific insights that cannot be achieved from a single sequence alone.

One important type of joint analysis is to identify features whose corresponding hypothesis tests are non-null in each sequence. These will be referred to here as simultaneous signals. To be precise, let $T_{id}$ be the test statistic corresponding to the $i$th feature in the $d$th sequence, for $i = 1, \ldots, n$ and $d = 1, \ldots, D$. Suppose each $T_{id}$ corresponds to a signal indicator $I_{id}$ that equals 1 if feature $i$ is truly significant in sequence $d$ and 0 otherwise. Let the signal configuration of the $i$th feature be represented by the true signal vector $\bI_i = (I_{i1}, \ldots, I_{iD})$, and define
\begin{equation}
  \label{eq:S}
  \begin{aligned}
    \mathcal{S}_D =\,& \left\{ (I_1, \ldots, I_D) \in \{0, 1\}^D: \sum_{d = 1}^D I_d = D \right\}.
  \end{aligned}
\end{equation}
Then feature $i$ is a simultaneous signal if $\bI_i \in \mathcal{S}_D$.

This paper studies the problem of identifying simultaneous signals, which arises frequently in many different contexts. In genetics it is frequently of interest to identify polymorphisms that are associated with multiple related conditions. These studies of what is termed genetic pleiotropy are common in recent research, for example in psychiatry \citep{cross2013genetic,cross2013identification,andreassen2013improved}. Similarly, it is useful to colocalize genotype-phenotype and genotype-gene expression associations to the same genetic variants, as loci that are simultaneously associated with both outcomes may contain important causal variants \citep{wallace2012statistical, fortune2015statistical, giambartolomei2018bayesian}. As another example, identifying findings that replicate across independent studies is a crucial component of reproducible research \citep{bogomolov2013discovering, heller2014deciding, heller2014replicability}. Finally, comparative genomics research aims to find genes whose orthologs are associated with similar phenotypes across multiple animal species, in hopes of finding evolutionarily conserved genomic programs \citep{rittschof2014neuromolecular, thompson2015comparative, saul2018cross}.

There has been a great deal of recent work on methods to control the false discovery rate when identifying these simultaneous signals. Let $\delta(T_{i1}, \ldots, T_{iD})$ be the result of applying a discovery procedure to the test statistics, such that $\delta(T_{i1}, \ldots, T_{iD}) = 1$ if the procedure declares $i$ to be a simultaneous signal and $\delta(T_{i1}, \ldots, T_{iD}) = 0$ otherwise. False discovery rate control methods aim to maximize the number of discovered simultaneous signals while maintaining the false discovery rate
\begin{equation}
  \label{eq:fdr}
  \textsc{fdr}(\delta)
  =
  E \left[
  \frac{
    \sum_{i \notin \mathcal{S}_D} \delta(T_{i1}, \ldots, T_{iD})
  }{
    \max\{1, \sum_{i = 1}^n \delta(T_{i1}, \ldots, T_{iD})\}
  }
  \right]
\end{equation}
to be at most $\alpha$, for some prespecified $\alpha < 1$.

A simple approach is to use a standard procedure, like that of \citet{benjamini1995controlling}, to discover significant features separately in each of the $D$ sequences of test statistics and then to identify discoveries common to all sequences. \citet{bogomolov2018assessing} developed a modified version of this idea and proved that their procedure maintain false discovery rate control. Another common strategy is to summarize the pair of statistics for each feature into a single scalar statistic, for example by taking the maximum of the corresponding $p$-values \citep{phillips2014testing}. This reduces the problem to a single sequence of multiple tests, but it is unclear how to choose the best summary function. A more principled approach is to treat the sequences as a single sequence of multivariate test statistics $(T_{i1}, \ldots, T_{iD})$. In this framework, it has been shown that the local false discovery rate \citep{efron2010large} is the optimal scalar summary of the multivariate test statistics \citep{chi2008false, chung2014gpa, du2014single, heller2014replicability}. This can be difficult to calculate in practice, so instead \citet{chung2014gpa} assumes a parametric model and used the EM algorithm to estimate unknown parameters, \citet{chi2008false} proposes a Taylor expansion approximation, \citet{du2014single} uses a single-index model approximation, and \citet{heller2014replicability} employed an empirical Bayes approach.

All of these methods assume that the null distributions of the test statistics $T_{id}$ are known. Many assume that $p$-values are available, meaning that the nulls must be known exactly. Others estimate empirical null distributions, which still requires knowing the parametric families to which the nulls belongs \citep{schwartzman2008empirical}. However, in many important problems in genomics, information about the null distributions of the $T_{id}$ is not readily available, for at least three common reasons. First, small sample sizes can make it difficult to obtain the exact null distribution of standard test statistics \citep{yu2013shrinkage}. Second, complex test statistics can have intractable null distributions. For example, the null distribution of the SKAT statistic \citep{wu2011rare}, which tests the significance of a set of genetic variants, does not have a convenient closed form and in practice is computationally approximated. Finally, complex data types can give rise to null distributions that are difficult to model or characterize. For example, data from ChIP-seq experiments \citep{park2009chip} are used to identify regions of the genome where transcription factors are found to bind, but the number, size, and locations of these regions are not predetermined. This makes accurate quantification of the statistical significance of the identified regions very difficult \citep{chitpin2018recap}.

To date, relatively little work has considered false discovery rate control when null distributions are not completely known. Some results are available given a single sequence of test statistics. Knockoff filters \citep{barber2015controlling, barber2016knockoff, arias2017distribution, candes2018panning} assume only that the null distributions are identical and symmetric, and $p$-filters \citep{barber2017p, ramdas2017unified} assume only that the test statistics can be converted to random variables between 0 and 1 that are stochastically larger than a uniformly distributed random variable. Resampling-based procedures, such as that of \citet{yekutieli1999resampling}, do not require known null distributions, but can only be used if the raw data are available. This may not be true for some applications, such as in genetics, where it is common that only test statistics are easily accessible. In contrast to the single sequence case, results are lacking when there are two or more test statistic sequences of interest. Nonparametric methods for detecting the presence of simultaneous signals have been proposed \citep{zhao2017sparse, zhao2017optimal}, but methods for identifying them when null distributions are unknown do not appear to exist.

This paper develops a new nonparametric method for false discovery rate control when discovering simultaneous signals with unknown null distributions. Section \ref{sec:proposed} describes the proposed procedure and shows that it can asymptotically control the false discovery rate at the nominal level under certain conditions. Section \ref{sec:alt} discusses an alternative procedure, originally proposed in an earlier preprint of this paper, that has more power but requires much more restrictive conditions. Section \ref{sec:sims} illustrates the performance of the proposed method in simulations, and Section \ref{sec:data} applies it to a simultaneous signal identification problem with unknown null distributions that was encountered when attempting to identify mouse genes that were both differentially expressed and whose neighboring chromatin was differentially accessible \citep{saul2017transcriptional}. Section \ref{sec:disc} concludes with a discussion, and proofs of all technical results can be found in the Appendix.

\section{\label{sec:proposed}Proposed procedure}
\subsection{\label{sec:2}Two sequences of test statistics}
For clarity of exposition, the proposed method is first introduced assuming that only two sequences of test statistics $T_{id}$ are observed, $i = 1, \ldots, n$ and $d = 1, 2$. Section \ref{sec:>=2} describes an extension to any number of sequences. Throughout the paper, the test statistics will be modeled as following
\begin{equation}
  \label{eq:model}
  T_{id} \mid I_{id} = 0 \sim F_{id}^0(t_d) = 1 - S_{id}^0(t_d),
  \quad
  T_{id} \mid I_{id} = 1 \sim F_{id}^1(t_d) = 1 - S_{id}^1(t_d),
\end{equation}
where $F_{id}^0(t_d)$ and $F_{id}^1(t_d)$ denote the null and alternative distributions for $T_{id}$ and $S_{id}^0(t_d)$ and $S_{id}^1(t_d)$ denote the corresponding survival functions. The $D$ sequences are assumed to be mutually independent, which models the setting where each sequence arises from an independent study. Finally, it will be assumed that the test statistics are two-tailed, in the sense that larger values of $T_{id}$ give more evidence against the null. This is formalized in the stochastic ordering condition of Assumption \ref{a:sto_ord}.

\begin{assumption}
  \label{a:sto_ord}
  For all $t_d$, $S^0_{id}(t_d) < S^1_{id}(t_d)$.
\end{assumption}

The overall strategy follows the framework of \citet{storey2004strong} for false discovery rate control in a single sequence of test statistics. The proposed procedure declares a features $i$ to be a simultaneous signal if
\begin{equation}
  \label{eq:region}
  (T_{i1}, T_{i2}) \in [t, \infty) \times [t, \infty)
\end{equation}
for an appropriately chosen threshold $t$. Many other rejection regions are possible. As mentioned in Section \ref{sec:intro}, it has been shown that the optimal rejection region is actually a level curve of the local false discovery rate. Nevertheless, \eqref{eq:region} is simple to implement and interpret, and is crucial to the nonparametric property of the proposed approach, as discussed at the end of this subsection.

One potential issue with \eqref{eq:region} is that using the same threshold for both $T_{i1}$ and $T_{i2}$ may not be appropriate if the test statistics are on different scales, in the sense that the null distributions $S_{i1}^0$ and $S_{i2}^0$ are not comparable. Indeed, it is perhaps more natural to consider regions $[t_1, \infty) \times [t_2, \infty)$ that allow different thresholds \citep{chi2008false, du2014single}. However, Section \ref{sec:alt} shows that this actually results in a procedure with unfavorable properties. Furthermore, the test statistics can be placed on the same scale by simply transforming the $T_{id}$ within each sequence to their corresponding ranks. Simulations in Section \ref{sec:sims} demonstrate that this strategy performs just as well as if the $T_{i1}$ and $T_{i2}$ were truly on the same scale.

The goal is to choose a threshold $t$ that discovers the most simultaneous signals while maintaining an acceptable false discovery rate. This requires estimating the false discovery proportion that would be attained by a particular threshold $t$. To motivate an estimator, suppose for now that the null and alternative distributions are the same across tests $i$, so that $S_{id}^0(t_d) = S_d^0(t_d)$ and $S_{id}^1(t_d) = S_d^1(t_d)$; this condition is much stronger than necessary and will be weakened in Assumption \ref{a:conv}. Define $\pi_{\bI}$ to be the proportion of features with true signal vector equal to $\bI$. Then under model \eqref{eq:model}, the expected proportion of false positives would equal
\[
\sum_{\bI \in \mathcal{S}_2^c} \pi_{\bI} S_1^{I_1}(t) S_2^{I_2}(t),
\]
where $\mathcal{S}_2^c = \{(0, 0), (0, 1), (1, 0)\}$ from \eqref{eq:S}. The following result shows that this expected proportion can be upper-bounded by the product of marginal survival functions.

\begin{proposition}
  \label{prop:bound}
  For proportions $\pi_{\bI}$ that satisfy $\sum_{\bI \in \{0, 1\}^2} \pi_{\bI} = 1$ and stochastically ordered survival functions $S_d^0(t_d) < S_d^1(t_d)$, $d = 1, 2$, define the marginal signal proportions $\pi_d = \sum_{\bI \in \{0, 1\}^2: I_d = 1} \pi_{\bI}$ and marginal survival functions $S_d(t_d) = (1 - \pi_d) S_d^0(t_d) + \pi_d S_d^1(t_d)$. Then
  \[
  \sum_{\bI \in \mathcal{S}_2^c} \pi_{\bI} S_1^{I_1}(t_1) S_2^{I_2}(t_2) \leq S_1(t_1) S_2(t_2)
  \]
  for any $t_1$ and $t_2$, with $\mathcal{S}_2$ defined in \eqref{eq:S}.
\end{proposition}

A reasonable estimate of an upper bound for the false discovery proportion that would be attained by the rejection region $[t, \infty) \times [t, \infty)$ is therefore
\begin{equation}
  \label{eq:fdp}
  \widehat{\textsc{fdp}}_\rho(t)
  =
  \frac{\hat{S}_1(t) \hat{S}_2(t) + \rho}{n^{-1} \vee \hat{G}(t, t)},
\end{equation}
where $\hat{S}_d(t_d) = n^{-1} \sum_{i = 1}^n I(T_{id} > t_d)$ are empirical marginal survival functions, $\hat{G}(t, t) = n^{-1} \sum_{i = 1}^n I(T_{i1} > t, T_{i2} > t)$ is the total proportion of rejected features, and $\rho$ is a positive constant that regularizes the asymptotic properties of the proposed procedure. An alternative to \eqref{eq:fdp} would be to define $\widehat{\textsc{fdp}}_\rho(t) = 0$ if $\hat{G}(t, t) = 0$, but \eqref{eq:fdp} is more convenient for proving asymptotic false discovery rate control. Because $T_{i1}$ and $T_{i2}$ are independent under model \eqref{eq:model}, it may seem that $\hat{G}(t, t)$ will always converge to $S_1(t) S_2(t)$. This does not happen because the $(T_{i1}, T_{i2})$ are not identically distributed across $i$. An apparent dependence is induced between the two sequences of test statistics by the different configurations of the true signal vectors $\bI_i$ for different $i$, and \eqref{eq:fdp} is closely related to testing for independence between the sequences of test statistics.

The proposed discovery procedure is therefore defined as
\begin{equation}
  \label{eq:proposed}
  \begin{aligned}
    \hat{\delta}_\rho(T_{i1}, T_{i2}) &= I(T_{i1} \geq \hat{t}_\rho, T_{i2} \geq \hat{t}_\rho), \\
    \hat{t}_\rho &= \inf \left\{ t \in [0, \infty) : \frac{\hat{S}_1(t) \hat{S}_2(t) + \rho}{n^{-1} \vee n^{-1} \sum_{i = 1}^n I(T_{i1} \geq t, T_{i2} \geq t)} \leq \alpha \right\},
  \end{aligned}
\end{equation}
for some desired false discovery rate $\alpha < 1$. Features with $\hat{\delta}_\rho(T_{i1}, T_{i2}) = 1$ are declared to be simultaneous signals. The threshold $\hat{t}_\rho$ maximizes the number of rejected features while maintaining $\widehat{\textsc{fdp}}_\rho(t) \leq \alpha$, which is a reasonable constraint because $\widehat{\textsc{fdp}}_\rho(t)$ is a conservative estimate of the true false discovery rate.

The proposed procedure  can be implemented without any knowledge of the null distributions $S_{id}^0$ beyond the stochastic ordering of Assumption \ref{a:sto_ord}. This stems from the rectangular shape of the rejection region \eqref{eq:region}, which gives rise to an expected false discovery proportion that can be upper bounded using only marginal survival functions, via Proposition \ref{prop:bound}. It is not clear whether there exist other rejection region shapes that can  endow a false discovery rate control procedure with this nonparametric property.

\subsection{\label{sec:theory}Theoretical properties}
The following assumption about the $S_{id}^0$ and $S_{id}^1$ will be required. Let $n_\bI$ denote the number of features with signal configuration $\bI$.

\begin{assumption}
  \label{a:conv}
  For sequences $d = 1, 2$ and $\theta = 0, 1$, there exist continuous functions $S_d^\theta(t)$ such that uniformly in $t_1$ and $t_2$,
  \begin{align*}
    \lim_{n \rightarrow \infty}
    \frac{1}{n_{(\theta, 0)} + n_{(\theta, 1)}} \sum_{i: I_{i2} = \theta} S_{i1}^\theta(t) = S_1^\theta(t_1),
    \quad
    \lim_{n \rightarrow \infty}
    \frac{1}{n_{(0, \theta)} + n_{(1, \theta)}} \sum_{i: I_{i1} = \theta} S_{i2}^\theta(t) = S_2^\theta(t_2),
  \end{align*}
  and for $\bI = (I_1, I_2) \in \{ (0, 0), (0, 1), (1, 0)\}$,
  \[
  \lim_{n \rightarrow \infty}
  \frac{1}{n_{\bI}} \sum_{i : \bI_i = \bI} S_{i1}^{I_1}(t_1) S_{i2}^{I_2}(t_2) = S_1^{I_1}(t_1) S_2^{I_2}(t_2).
  \]
  There also exists a continuous function $G^1(t_1, t_2)$ such that uniformly in $t_1$ and $t_2$,
  \begin{align*}
    \lim_{n \rightarrow \infty}
    \frac{1}{n_{(1, 1)}} \sum_{i: \bI_i = (1, 1)} S_{i1}^1(t_1) S_{i2}^1(t_2) = G^1(t_1, t_2).
  \end{align*}
  Finally, there exist proportions $\pi_\bI$ such that $n_\bI / n \rightarrow \pi_{\bI}$ for every $\bI \in \{0, 1\}^2$.
\end{assumption}

Assumption \ref{a:conv} is trivially satisfied when the $S_{id}^0(t_d)$ and $S_{id}^1(t_d)$ do not vary across $i$. Otherwise, the limiting $S_d^0(t_d)$ and $S_d^1(t_d)$ can be thought of as mixtures of the features-specific distributions. A similar assumption was also made in \citet{storey2004strong}. The second display in Assumption \ref{a:conv} requires that across each set of features that are not simultaneous signals, the $S_{i1}^0(t_1)$ and $S_{i2}^0(t_2)$ should be uncorrelated, in the sense that in the limit, the average of their inner product should equal the product of their averages.

\begin{theorem}
  \label{thm:fdr}
  Under Assumptions \ref{a:sto_ord} and \ref{a:conv}, the proposed procedure \eqref{eq:proposed} with $\rho > 0$ satisfies
  \[
  \limsup_{n \rightarrow \infty} \textsc{fdr}(\hat{\delta}_\rho) \leq \alpha,
  \]
  where $\textsc{fdr}(\hat{\delta}_\rho)$ is the true false discovery rate defined in \eqref{eq:fdr}.
\end{theorem}

Theorem \ref{thm:fdr} states that the proposed procedure can achieve asymptotic false discovery rate control. Finite-sample rather than asymptotic control would be ideal, and the proof of Theorem \ref{thm:fdr} shows that this would be possible if the marginal survival functions $S_d(t_d)$ in $\widehat{\textsc{fdr}}_\rho(t)$ \eqref{eq:fdp} were known rather than estimated. The condition that $\rho > 0$ is necessary for technical reasons, but the simulations in Section \ref{sec:sims} indicates that using $\rho = 0$ still gives extremely good performance in practice.

\subsection{\label{sec:>=2}More than two sequences of test statistics}
In some problems, the goal may be to discover features that are simultaneously significant across $D \geq 2$ sequences of test statistics. The proposed method can be extended to this setting by consider rejection regions of the form $[t, \infty)^D$ for a threshold $t$. Under model \eqref{eq:model} and Assumption \ref{a:conv}, the expected number of false positives discovered that would be discovered by this region equals $\sum_{\bI \in \mathcal{S}_D^c} \pi_{\bI}  \prod_{d = 1}^D S_d^{I_d}(t)$, which can be upper-bounded by marginal survival functions using the following generalization of Proposition \ref{prop:bound}.

\begin{proposition}
  \label{prop:bound>=2}
  For $D \geq 2$, for proportions $\pi_{\bI}$ that satisfy $\sum_{\bI \in \{0, 1\}^D} \pi_{\bI} = 1$,
  \[
  \sum_{\bI \in \mathcal{S}_D^c} \pi_{\bI}  \prod_{d = 1}^D S_d^{I_d}(t_d) \leq \sum_{d, d' \in 1,\ldots, D, d \ne d'} S_d(t_d) S_{d'}(t_{d'})
  \]
  for any $t_1, \ldots, t_D$, with marginal survival functions $S_d(t_d)$ defined as in Proposition \eqref{prop:bound} and $\mathcal{S}_D$ defined in \eqref{eq:S}.
\end{proposition}

Following the reasoning in Section \ref{sec:2}, Proposition \ref{prop:bound>=2} therefore motivates the following discovery procedure for any number $D \geq 2$ of sequences:
\begin{equation}
  \label{eq:proposed>=2}
  \begin{aligned}
    \hat{\delta}_\rho(T_{i1}, \ldots, T_{iD}) &= I(T_{i1} \geq \hat{t}_\rho, \ldots, T_{iD} \geq \hat{t}_\rho), \\
    \hat{t}_\rho &= \inf \left\{ t \in [0, \infty) : \frac{\sum_{d, d' \in 1, \ldots, D, d \ne d'} \hat{S}_d(t) \hat{S}_{d'}(t) + \rho}{n^{-1} \vee n^{-1} \sum_{i = 1}^n I(T_{i1} \geq t, \ldots, T_{iD} \geq t)} \leq \alpha \right\},
  \end{aligned}
\end{equation}
and features with $\hat{\delta}_\rho(T_{i1}, \ldots, T_{iD}) = 1$ are declared as simultaneous signals across all $D$ sequences. It is straightforward to extend the proof of Theorem \ref{thm:fdr} to this generalized procedure.

Though \eqref{eq:proposed>=2} can asymptotically control the false discovery rate, it can be highly conservative, meaning that it may not make many discoveries. The reason is that Proposition \ref{prop:bound>=2} does not provide a very tight bound on the expected number of false positives. The original inequality in Proposition \ref{prop:bound} for two sequences of test statistics is related to the close connection between simultaneous signals and dependence between the signals, as discussed in Section \ref{sec:2}. However, this connection disappears when there are more than two sequences of tests, as dependence between the sequences no longer implies the existence of simultaneous signals. Thus inequalities like Propositions \ref{prop:bound} and \ref{prop:bound>=2} may not be the optimal approach in this case, and further work is necessary to design a more powerful nonparametric discovery procedure for more than two sequences.

\section{\label{sec:alt}Alternative procedures}
\subsection{\label{sec:alt_method}Methodology and theoretical properties}
As discussed in Section \ref{sec:2}, a more natural alternative to rejection region \eqref{eq:region} is the rectangle $[t_1, \infty) \times [t_2, \infty)$, which allows a different threshold for each sequence of test statistics. Applying Proposition \ref{prop:bound} suggests the new false discovery proportion bound
\begin{equation}
  \label{eq:alt_fdp}
  \widetilde{\textsc{fdp}}_\rho(t_1, t_2) = \frac{\hat{S}_1(t_1) \hat{S}_2(t_2) + \rho}{n^{-1} \vee \hat{G}(t_1, t_2)},
\end{equation}
which can be shown to be an asymptotically uniformly conservative estimate of false discovery rate incurred by the rejection region.
    
\begin{theorem}
  \label{thm:alt_conservative}
  For any discovery procedure of the form $\delta(T_{i1}, T_{i2}) = I(T_{i1} \geq t_1, T_{i2} \geq t_2)$, under Assumptions \ref{a:sto_ord} and \ref{a:conv},
  \[
  \lim_{p\rightarrow\infty}
  \inf_{t_1 \leq \eta_1, t_2 \leq \eta_2}
  \left\{\widetilde{\textsc{fdp}}_\rho(t_1,t_2) - \textsc{fdr}(\delta)\right\}
  \geq 0
  \]
  almost surely, for fixed $\eta_1,\eta_2 < \infty$.
\end{theorem}

The false discovery proportion bound \eqref{eq:alt_fdp} leads to the following procedure, originally proposed in an earlier preprint of this paper:
\begin{equation}
  \label{eq:alt}
  \begin{aligned}
    \tilde{\delta}_\rho(T_{i1}, T_{i2}) &= I(T_{i1} \geq \hat{t}_{\rho 1}, T_{i2} \geq \hat{t}_{\rho 2}), \\
    (\hat{t}_{\rho, 1}, \hat{t}_{\rho 2}) &= \argmax_{(t_1, t_2) \in \Pi} \hat{G}(t_1, t_2) \mbox{ subject to } \widetilde{\textsc{fdp}}_\rho(t_1, t_2) \leq \alpha,
  \end{aligned}
\end{equation}
where the set $\Pi=\{(\infty,\infty)\} \cup \{(T_{i1}, T_{i'2}) : 1 \leq i, i' \leq n\}$ is the union of the point $(\infty,\infty)$ along with the Cartesian product of the two sequences of observed test statistics. The $\hat{t}_{\rho 1}$ and $\hat{t}_{\rho 2}$ are chosen to maximize $\hat{G}(t_1, t_2)$, which is equivalent to maximizing the number of rejected features, subject to controlling the estimated false discovery rate bound. Under certain conditions, $\tilde{\delta}$ achieve asymptotic false discovery rate control. Let $\textsc{fdp}_\rho(t_1,t_2)$ denote the pointwise limit of $\widetilde{\textsc{fdp}}_\rho(t_1, t_2)$.

\begin{theorem}
  \label{thm:alt_fdr}
  Under Assumptions \ref{a:sto_ord} and \ref{a:conv}, if there exist $t'_1,t'_2 < \infty$ such that $\textsc{fdp}_\rho(t'_1,t'_2) < \alpha$, the alternative procedure \eqref{eq:alt} satisfies
  \[
  \limsup_{n \rightarrow \infty} \textsc{fdr}(\tilde{\delta}_\rho) \leq \alpha.
  \]
\end{theorem}

\subsection{\label{sec:alt_issues}Issues}
\begin{figure}
  \centering
  \includegraphics[scale = 0.75]{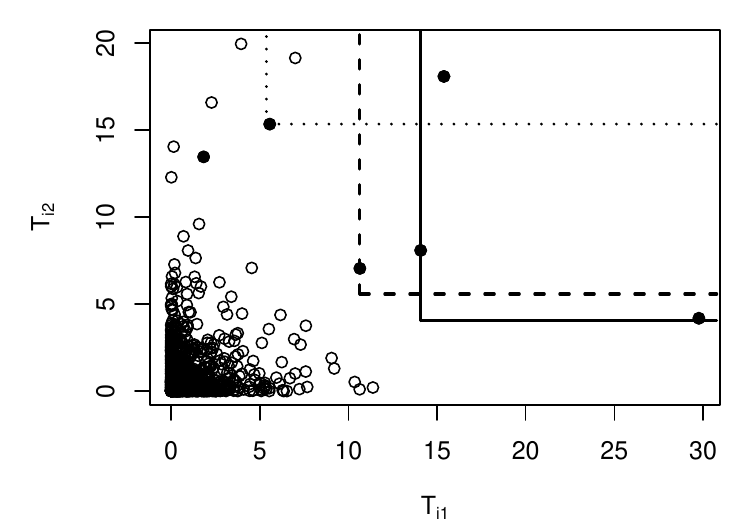}
  \caption{\label{fig:nonunique}The solid, dashed, and dotted lines demarcate three distinct rejection regions, each of which maximizes the number of rejections while satisfying $\widetilde{\textsc{fdp}}_\rho(t_1, t_2) \leq 0.05$. Filled circles denote true simultaneous signals.}
\end{figure}

Procedure \eqref{eq:alt} is evidently more flexible than the proposed procedure \eqref{eq:proposed}, as it allows two different thresholds. However, it suffers from two issues. First, the $\hat{t}_{\rho 1}$ and $\hat{t}_{\rho 2}$ defined in \eqref{eq:alt} are not unique. One reason is that $\hat{G}(t_1, t_2)$ and $\widetilde{\textsc{fdp}}_\rho(t_1, t_2)$ are piecewise-constant functions. Another is that there can exist multiple distinct rejection regions that maximize $\hat{G}(t_1, t_2)$. Figure \ref{fig:nonunique} illustrates an example where $n = 1,000$, $\pi_{(0, 0)} = 0.985$, $\pi_{(1, 0)} = \pi_{(0, 1)} = \pi_{(1, 1)} = 0.005$, the null $T_{id} \sim \chi^2_1$, and the non-null $T_{id} \sim \chi^2_1(9)$. Each of the rectangular rejection regions at the $\alpha = 0.05$ level rejects a different set of three features. While Theorem \ref{thm:alt_fdr} applies to any $\hat{t}_{\rho 1}$ and $\hat{t}_{\rho 2}$ that satisfies \eqref{eq:alt}, in practice it may not be clear which to choose.

The second issue concerns the requirement that there exist $t'_1$ and $t'_2$ such that the pointwise limit $\textsc{fdp}_\rho(t'_1, t'_2) \leq \alpha$. This is needed to prove the existence of finite upper-bounds on all feasible $\hat{t}_{\rho 1}$ and $\hat{t}_{\rho 2}$, so that the uniformity of the result of Theorem \ref{thm:alt_conservative} can be applied, but is very stringent. To see this, assume for simplicity that the null and alternative distributions do not vary across features. Following the proof of Proposition \ref{prop:bound}, it can be shown that
\begin{align*}
  \textsc{fdp}_\rho(t_1, t_2)
  =
  \frac{S_1(t_1) S_2(t_2) + \rho}{S_1(t_1) S_2(t_2) + (\pi_{(1, 1)} - \pi_1\pi_2) \{S^1_1(t_1) - S^0_1(t_1)\}\{S^1_2(t_2) - S^0_2(t_2)\}},
\end{align*}
where $\pi_1$ and $\pi_2$ are the marginal proportions of non-null features in each of the sequences, defined in Proposition \ref{prop:bound}, and $\pi_{(1, 1)}$ is the proportion of simultaneous signals, defined in Assumption \ref{a:conv}. The existence condition on $(t'_1, t'_2)$ now requires that $\pi_{(1, 1)} > \pi_1 \pi_2$, otherwise $\textsc{fdp}_\rho(t_1,t_2) \geq 1$ for all $(t_1, t_2)$. In other words, Theorem \ref{thm:alt_fdr} cannot guarantee that procedure $\tilde{\delta}$ will control the false discovery rate unless the proportion of simultaneous signals is large enough, which can be difficult to verify.

These issues restrict the practical application of the alternative procedure \eqref{eq:alt}. Furthermore, Section \ref{sec:2} provided a rank transformation strategy that can place the $T_{i1}$ and $T_{i2}$ on the same scale, obviating the need for a different threshold for each sequence of test statistics. Therefore, this alternative procedure is not pursued in the remainder of this paper.

\section{\label{sec:sims}Simulations}
\subsection{\label{sec:sims_scale}Performance of rank transformation}
As mentioned in Section \ref{sec:proposed}, the rejection region of the proposed method uses the same threshold for each sequence of test statistics. This will not work well if the null distributions of the different sequences are of different scales, but Section \ref{sec:2} describes a procedure to rescale the $T_{id}$ by transforming them to their corresponding ranks within each sequence.

The simulations in this section explore the effectiveness of this transformation for $D = 2$ sequences. In sequence 1, $T_{i1}$ were drawn independently from N$(\mu_{i1}, 1)$, where $\mu_{i1} = 0$ if the corresponding signal indicator $I_{i1} = 0$ and otherwise was drawn from N$(5, 1)$ and fixed across all replications. In sequence 2, $T_{i2}$ were drawn independently from N$(\mu_{i2}, 4)$, with the $\mu_{i2}$ independently generated similar to the $\mu_{i1}$. The null distribution of $T_{i2}$ is therefore different from that of $T_{i1}$. The larger variance of $T_{i2}$ means that larger values $T_{i2}$ can still correspond to null features.

The proposed discovery procedure \eqref{eq:proposed}, with the regularization parameter $\rho$ set to zero, was applied in three ways. First, knowledge of the true null distributions was used to calculate two-tailed $p$-values $P_{i1}$ and $P_{i2}$, and the proposed method was applied without rank transformation to $-\log_{10} P_{id}$. These results represent the performance of the proposed method when the test statistics are on comparable scales. The method was also directly applied to $T_{id}^2$ to illustrate the impact of having different scales. Finally, the method was applied to rank-transformed $T_{id}^2$.

\begin{figure}
  \centering
  \includegraphics[scale = 0.7]{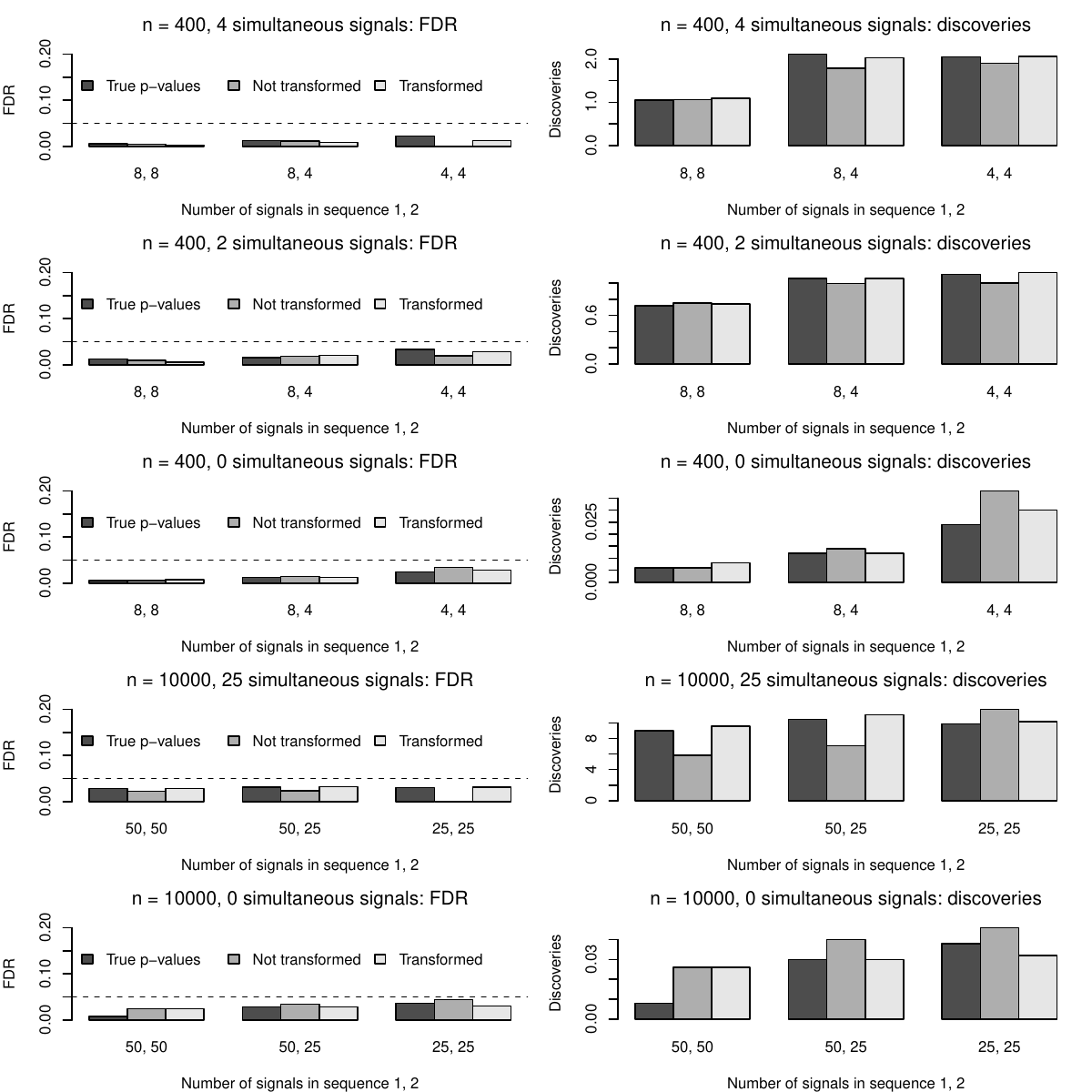}
  \caption{\label{fig:scale}Simulations showing the impact of rank-transformation in $D = 2$ sequences of test statistics. Horizontal dashed lines mark the nominal 0.05 false discovery rate level. True $p$-values: the proposed method \eqref{eq:proposed} applied to $-\log_{10} P_{id}$. Not transformed: \eqref{eq:proposed} applied to $T_{id}^2$. Transformed: \eqref{eq:proposed} applied to rank-transformed $T_{id}^2$.}
\end{figure}

Figure \ref{fig:scale} reports the results after 500 replications. All implementations of the proposed procedure maintained the false discovery rate at the nominal $\alpha = 0.05$ level. Using the true $p$-values identified the most simultaneous signals. Applying the procedure without transforming performed worse in many cases, and using rank-transformed $T_{id}^2$ performed nearly as well using the true $p$-values. Rank transformation thus appears to be an easy way to recover the optimal performance of the proposed method, and is always used in the remainder of this paper.

\subsection{\label{sec:sim_comparison}Comparison to existing methods}
The proposed procedure, with $\rho = 0$ and rank-transformed $T_{id}$, was compared to three existing methods described in Section \ref{sec:intro}. The method of \citet{chung2014gpa} imposes parametric assumptions on $p$-values $P_{id}$, calculated from the $T_{id}$, under the alternative distribution. The empirical Bayes method of \citet{heller2014replicability} estimates an empirical null distribution of $z$-scores calculated from the $T_{id}$. Finally, the method of \citet{bogomolov2018assessing} is based on first selecting promising features from each sequence based on the $P_{id}$. These existing approaches all require calculating either $p$-values or $z$-scores and therefore require knowledge of the true null distributions of the $T_{id}$.

\begin{figure}
  \centering
  \includegraphics[scale = 0.7]{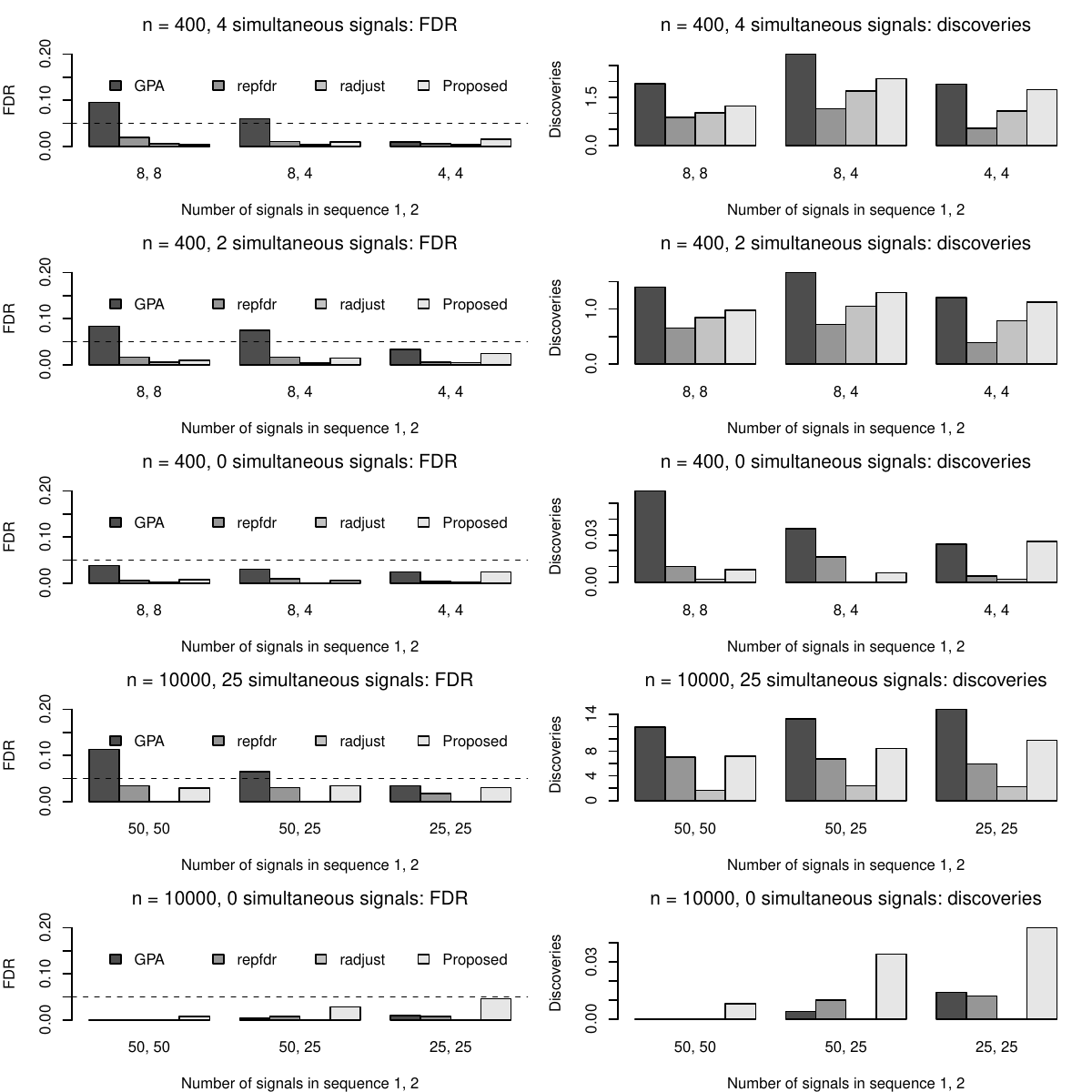}
  \caption{\label{fig:norm}Example 1. Simulations of $D = 2$ sequences of test statistics with known null distributions. Horizontal dashed lines mark the nominal 0.05 false discovery rate level. GPA: method of \citet{chung2014gpa}; repfdr: method of \citet{heller2014replicability}; radjust: method of \citet{bogomolov2018assessing}; Proposed: proposed approach \eqref{eq:proposed}.}
\end{figure}

\textit{Example 1.} These methods were first applied to a typical setting with $D = 2$ sequences and known null distributions. The $T_{id}$ were independently generated from N$(\mu_{id}, 1)$, with $\mu_{id} = 0$ if $I_{id} = 0$ and otherwise drawn from N$(3, 1)$ and fixed across replications. The proposed procedure was applied to the $T_{id}^2$. Figure \ref{fig:norm} shows that the procedure of \citet{chung2014gpa} was most powerful but in some cases could not maintain the nominal false discovery rate. Among the remaining methods, the proposed procedure actually had the highest power in many of the simulation settings, while always maintaining the false discovery rate at the nominal level. This is perhaps because existing methods require good estimates of the proportions of non-null signals, which are difficult to obtain in the highly sparse scenarios simulated here. The proposed method avoids this estimation problem. On the other hand, \citet{bogomolov2018assessing} show that the proposed method can have low power when the sequences are not very sparse. In this case, the bound in Proposition \ref{prop:bound} on the expected number of false positives is not very tight, and algorithms like that of \citet{chung2014gpa} have been shown to perform extremely well.

\textit{Example 2.} The methods were next applied to $D = 2$ sequences with $n = 1,000$ independent features, where the null distributions were unknown. To generate each $T_{id}$, $z$-scores $(Z_{id1}, \ldots, Z_{id10})$ were first generated from N$(\mu_{id}, \Sigma_{id})$, where $\mu_{id} = (0, \dots, 0)$ if $I_{id} = 1$ and otherwise was drawn from N$(2, 1)$ and fixed across replications. Each $\Sigma_{id}$ was equal to the empirical correlation matrix of a different set of 10 genes selected from a gene expression study of multiple myeloma, obtained from \citet{shi2010microarray}. Next, these $z$-scores were converted to correlated $p$-values $(P_{id1}, \ldots, P_{id10})$, and finally $T_{id} = -2 \sum_{j = 1}^{10} \log P_{idj}$. This setting models applications involving test statistics based on groups of genomic features, which are frequently correlated. The null distribution of each $T_{id}$ is complicated and in practice would not be known, as they depend on the unknown correlations because genomic features.

\begin{figure}
  \centering
  \includegraphics[scale = 0.7]{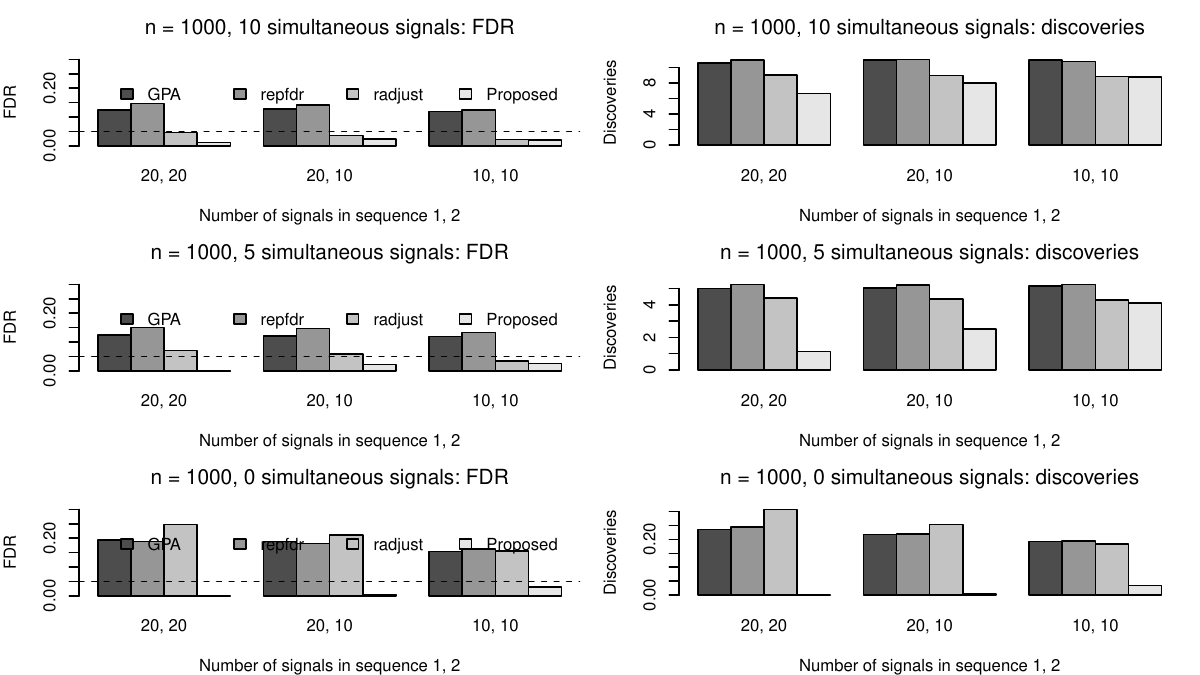}
  \caption{\label{fig:group}Example 2. Simulations of $D = 2$ sequences of test statistics with unknown null distributions. Horizontal dashed lines mark the nominal 0.05 false discovery rate level. GPA: method of \citet{chung2014gpa}; repfdr: method of \citet{heller2014replicability}; radjust: method of \citet{bogomolov2018assessing}; Proposed: proposed approach \eqref{eq:proposed}.}
\end{figure}

Figure \ref{fig:group} reports the results after 500 replications. The proposed procedure was applied to the $T_{id}$. The methods of \citet{chung2014gpa}, \citet{heller2014replicability}, and \citet{bogomolov2018assessing} require known null distributions. Here they were implemented assuming that the $T_{id}$ followed $\chi^2_{20}$ under the null, which would only be correct if the genes were independent. In practice, it would be inappropriate to use these methods if the nulls were unknown, and here they are only included to illustrate the consequences of misspecifying the null. Indeed, the simulations show that they do not maintain the false discovery rate at the nominal level. In contrast, the proposed method always maintains the nominal level and can have very good power.

\textit{Example 3.} The generalized discovery procedure in Section \ref{sec:>=2} was applied to $D = 3$ sequences with $n = 10000$ features. In sequences $d = 1, 2$, $T_{id}$ were independently generated from N$(\mu_{id}, 1)$, where $\mu_{id} = 0$ when $I_{id} = 0$ and otherwise was drawn from N$(5, 1)$ and fixed across replications. In the third sequence, $T_{i3}$ was generated from a complicated distribution meant to model the ChIP-seq data studied in Section \ref{sec:data}. First, $\lambda_{i1} = \lambda_{i2}$ were drawn from N$(100, 5)$ when $I_{i3} = 0$ and then fixed across replications. These model population average ChIP-seq peak heights at genomic location $i$ under experimental and control conditions, respectively, that are equal under the null hypothesis. When $I_{i3} = 1$, $\lambda_{i1}$ and $\lambda_{i2}$ were independently drawn from Exp$(0.001)$, modeling differences in average peak heights between the experimental conditions under the alternative hypothesis. Next, $O_{il}$ for $l = 1, 2$ were generated from Poisson($\lambda_{il}$), modeling observed ChIP-seq peak counts. Finally, $T_{i3} = \vert \log (O_{i1} / O_{i2}) \vert$, and will tend to be larger when $I_{i3} = 1$ because $\lambda_{i1} \ne \lambda_{i2}$.

\begin{table}
  \centering
  \begin{tabular}{r|cc|cc}
    & \multicolumn{4}{c}{Number of simultaneous signals} \\
    Number of signals & \multicolumn{2}{c}{FDR} & \multicolumn{2}{c}{Discoveries} \\
    in sequences 1, 2 & 25 & 0 & 25 & 0 \\
    \hline
    50, 50 & 0.001 & 0.000 & 16.486 & 0.000 \\
    50, 25 & 0.003 & 0.000 & 11.696 & 0.000 \\
    25, 25 & 0.000 & 0.000 & 17.848 & 0.000 \\
    \hline
  \end{tabular}
  \caption{\label{tab:multi}Example 3. Simulations of $D = 3$ sequences of test statistics with unknown null distributions, for the proposed approach \eqref{eq:proposed} with a nominal 0.05 false discovery rate level.}
\end{table}

Table \ref{tab:multi} reports the results over 500 replications.  The proposed procedure was applied to $(T_{i1}^2, T_{i2}^2, T_{i3})$. The null distributions of the $T_{i3}$ are complicated, making it difficult to apply any other existing methods. However, the proposed nonparametric procedure can be directly employed, and results show that it maintained the nominal false discovery rate while still being able to detect a significant proportion of the true simultaneous signals. That the attained false discovery rates are much lower than the nominal 0.05 indicates that the procedure is conservative, as discussed in Section \ref{sec:>=2}.

\begin{figure}
  \centering
  \includegraphics[scale = 0.7]{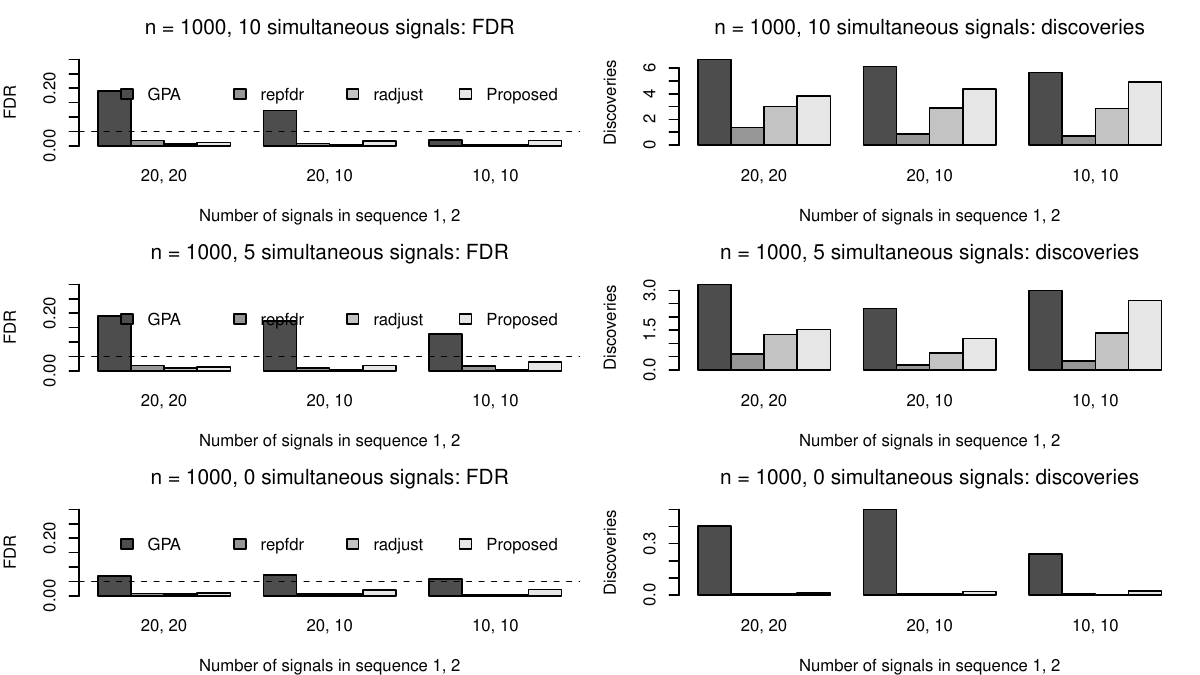}
  \caption{\label{fig:dep}Example 4. Simulations of $D = 2$ sequences of dependent test statistics with known null distributions. Horizontal dashed lines mark the nominal 0.05 false discovery rate level. GPA: method of \citet{chung2014gpa}; repfdr: method of \citet{heller2014replicability}; radjust: method of \citet{bogomolov2018assessing}; Proposed: proposed approach \eqref{eq:proposed}.}
\end{figure}

\textit{Example 4.} The proposed discovery procedure, along with most existing methods to simultaneous signal discovery, was developed assuming independence between the test statistics within each sequence. These simulations test its robustness to violations of this assumption for $D = 2$ sequences with $n = 1,000$ features. Each $T_{id}$ was marginally generated from $N(\mu_{id}, 1)$, where $\mu_{id} = 0$ when $I_{id} = 0$ and otherwise was drawn from N$(3, 1)$. Within each sequence, the correlation matrix of the $T_{id}$ was set equal to the empirical correlation matrix of 1,000 genes chosen from a gene expression study of multiple myeloma \citep{shi2010microarray}. Figure \ref{fig:dep} displays the results of 500 replications. The proposed method applied to $T_{id}^2$ and was able to maintain the false discovery rate at the nominal level while exhibiting excellent power. These results lend more confidence to using the method in practical applications.

\section{\label{sec:data}Data analysis}
The field of sociogenomics studies molecular correlates of social behavior \citep{robinson2005sociogenomics}. \citet{saul2017transcriptional} studied the transcriptomic response to social challenge in mice that were exposed to intruder mice introduced to their cages. At 30, 60, and 120 minutes after intruder removal, they collected RNA-seq data from the amygdala, frontal cortex, and hypothalamus in order to determine which genes were differentially expressed between mice exposed to the intruder and mice exposed to a nonsocial control condition. They also collected ChIP-seq H3K27ac data at 30 and 120 minutes, to identify regions of chromatin that were differentially accessible between experimental and control mice. These data are available from the Gene Expression Omnibus under accession number GSE80345.

This section analyzes these data to find mouse genes that are both differentially expressed and next to differentially accessible regions of chromatin. Integrating these pieces of evidence can identify genes whose expression changes may be directly caused by differential binding of transcription factors to nearby regions of DNA \citep{saul2017transcriptional}. This analysis can be cast as a simultaneous signal detection problem. Each mouse gene constitutes a genomic feature $i$, which can be associated with both a differential expression test statistic $T_{i1}$ and a test statistic $T_{i2}$ for the differential accessibility of a neighboring region of chromatin. The goal is to identify genes whose $T_{i1}$ and $T_{i2}$ are simultaneously non-null.

Following \citet{saul2017transcriptional}, the $T_{i1}$ were standard $z$-scores obtained using the edgeR software package \citep{robinson2010edger}. Defining $T_{i2}$ was more involved. Methods exist for calculating differential accessibility test statistics for genomic regions using ChIP-seq data \citep{zhang2008model, heinz2010simple, shen2013diffreps}, but these first identify regions of interest from the same data that the test statistics come from. This makes accurate $p$-values difficult to calculate \citep{chitpin2018recap}. This analysis takes a simple approach and by defining $T_{i2} = \vert \log (O_{i1} / O_{i2}) \vert$, where $O_{i1}$ and $O_{i2}$ were the observed number of H3K27ac reads, in the experimental and control sample respectively, within 100 kb up- and down-stream of the $i$th gene.

\begin{table}
  \centering
  \begin{tabular}{cc|cc|cc}
    \multicolumn{2}{c}{Amygdala} & \multicolumn{2}{c}{Frontal cortex} & \multicolumn{2}{c}{Hypothalamus} \\
    30 min & 120 min & 30 min & 120 min & 30 min & 120 min \\
    \hline
    & Klk6 & Nts & & Ai606473 & Lhx9 \\
    &&&& Foxg1 &\\
    &&&& Gpr88 &\\
    &&&& Meis2 & \\
    &&&& Penk & \\
    &&&& Slc5a7 &
  \end{tabular}
  \caption{\label{tab:data}Mouse genes found to be both differentially expressed and next to differentially accessible chromatin at a nominal false discovery rate of 0.1.}
\end{table}

The null distribution of $T_{i2}$ is highly nontrivial, and the proposed method \eqref{eq:proposed} is the only existing false discovery rate control procedures that can be used without knowledge of this null. Table \ref{tab:data} presents the genes identified at a nominal false discovery rate of 0.1. It indicates that the hypothalamus is the most transcriptionally responsive to social challenge, particularly at 30 minutes. A number of these genes have been previously implicated in mouse behavior. For example, mice without Gpr88 and Penk have been shown to exhibit low anxiety and resistance to mild stress \citep{melo2014enkephalin, meirsman2016mice}, and Foxg1 was highlighted in \citet{saul2017transcriptional} as providing evidence for the role of neuropeptide signaling and neuron differentiation. These findings raise novel mechanistic hypotheses about the molecular response to social challenge.

\section{\label{sec:disc}Discussion}
Most of this paper has assumed that the test statistics are independent across features. In the single-sequence false discovery rate control problem with dependent test statistics, an important step is to estimate an empirical null distribution for the test statistics rather than using the theoretical null \citep{efron2007correlation,efron2010correlated,schwartzman2012comment}. The nonparametric false discovery proportion bound \eqref{eq:fdp} already uses empirical distribution estimates, so the proposed procedure may also be able to control the false discovery proportion under dependence. This seems to be corroborated by the simulation results in Figure \ref{fig:dep}, but more work is required to fully characterize the behavior of the proposed method with dependent test statistics.

In some cases the two sequences of $p$-values are not of equal importance, as in replicability analysis \citep{bogomolov2013discovering, heller2014deciding, heller2014replicability, bogomolov2018assessing}, which distinguishes between a primary versus a follow-up study. The proposed method makes no such distinction, but could be potentially be modified. Suppose for two sequences that sequence 2 were of greater interest. Then the rejection region could be defined as $[t, \infty) \times [ct, \infty)$ for some fixed constant $0 < c < 1$. This may allow weaker signals to be captured from the more important study.

In simulations, the proposed procedure could sometimes be more powerful than existing simultaneous signal detection procedures even when the null distributions were known. As discussed in Section \ref{sec:sim_comparison}, this may be a consequence of the highly sparse sequences used in the simulations, for other methods perform much better when the null distributions are known and the signals are not too sparse \citep{chung2014gpa, bogomolov2018assessing}. It would be interesting to pursue nonparametric detection methods with good power in this moderately sparse regime.

\section*{Acknowledgments}
The authors would like to thank Dr. Michael Saul for providing the data from the mouse sociogenomic experiment. The work of Dave Zhao was funded in part by NSF grant DMS-1613005.

\bibliographystyle{abbrvnat}
\bibliography{refs}

\appendix

\section{Proof of Proposition \ref{prop:bound}}
By definition,
\begin{align*}
  S_1(t_1) S_2(t_2)
  =\,&
  (1-\pi_1) (1-\pi_2) S^0_1(t_1) S^0_2(t_2)
  +
  (1-\pi_1) \pi_2 S^0_1(t_1) S^1_2(t_2)
  \,+
  \\
  &
  \pi_1 (1-\pi_2) S^1_1(t_1) S^0_2(t_2)
  +
  \pi_1 \pi_2 S^1_1(t_1) S^1_2(t_2)
  \\
  =\,&
  (1 - \pi_1 - \pi_2) S^0_1(t_1) S^0_2(t_2)
  \,+
  \\
  &
  \pi_1 \pi_2 \{S^0_1(t_1) S^0_2(t_2) + S^1_1(t_1) S^1_2(t_2) - S^0_1(t_1) S^1_2(t_2) - S^1_1(t_1) S^0_2(t_2)\}
  \,+
  \\
  &
  \pi_2 S^0_1(t_1) S^1_2(t_2) + \pi_1 S^1_1(t_1) S^0_2(t_2).
\end{align*}
Combined with the definitions of the marginal signal proportions $\pi_d$, the above expression becomes
\begin{align*}
  S_1(t_1) S_2(t_2)
  =\,&
  (\pi_{00} - \pi_{11}) S^0_1(t_1) S^0_2(t_2)
  +
  (\pi_{01} + \pi_{11}) S^0_1(t_1) S^1_2(t_2)
  +
  (\pi_{10} + \pi_{11}) S^1_1(t_1) S^0_2(t_2)
  \,+
  \\
  &
  \pi_1 \pi_2 \{S^1_1(t_1) - S^0_1(t_1)\}\{S^1_2(t_2) - S^0_2(t_2)\}
  \\
  =\,&
  \pi_{00} S^0_1(t_1) S^0_2(t_2)
  +
  \pi_{01} S^0_1(t_1) S^1_2(t_2)
  +
  \pi_{10} S^1_1(t_1) S^0_2(t_2)
  \,+
  \\
  &
  \pi_{11} [S^0_1(t_1) S^1_2(t_2) + \{S^1_1(t_1) - S^0_1(t_1)\} S^0_2(t_2)]
  \,+
  \\
  &
  \pi_1 \pi_2 \{S^1_1(t_1) - S^0_1(t_1)\}\{S^1_2(t_2) - S^0_2(t_2)\}.
\end{align*}
Since $S^1_1(t_1) > S^0_1(t_1)$ by the stochastic ordering in Assumption \ref{a:sto_ord}, the middle term on the right-hand side of the last equality is always positive.

\section{Proof of Theorem \ref{thm:fdr}}
Define $\mathcal{R}(t) = \{i : I(T_{i1} \geq t, T_{i2} \geq t)\}$ to be the set of features rejected at threshold $t$. Then
\[
\textsc{fdr}(\hat{\delta}_\rho)
=
\sum_{i: \bI_i \in \mathcal{S}_2^c \cap \mathcal{R}(\hat{t}_\rho)}
E \frac{I(T_{i1} \geq \hat{t}_\rho, T_{i2} \geq \hat{t}_\rho)}{1 \vee \sum_{i = 1}^n I(T_{i1} \geq \hat{t}_\rho, T_{i2} \geq \hat{t}_\rho)},
\]
with $\mathcal{S}_2 = \{(1, 1)\}$ defined as in \eqref{eq:S}. Since $\hat{t}_\rho$ satisfies $\widehat{\textsc{fdp}}_\rho(\hat{t}_\rho) \leq \alpha$,
\[
\frac{\hat{S}_1(\hat{t}_\rho) \hat{S}_2(\hat{t}_\rho) + \rho}{n^{-1} \vee n^{-1} \sum_{i = 1}^n I(T_{i1} \geq \hat{t}_\rho, T_{i2} \geq \hat{t}_\rho)} \leq \alpha
\]
by the definition of $\widehat{\textsc{fdp}}_\rho(\hat{t}_\rho)$ in \eqref{eq:fdp}. Therefore,
\[
\textsc{fdr}(\hat{\delta}_\rho)
\leq
\frac{\alpha}{n}
\sum_{i: \bI_i \in \mathcal{S}_2^c \cap \mathcal{R}(\hat{t}_\rho)}
E \frac{I(T_{i1} \geq \hat{t}_\rho, T_{i2} \geq \hat{t}_\rho)}{\hat{S}_1(\hat{t}_\rho) \hat{S}_2(\hat{t}_\rho) + \rho}.
\]

Analogous to the proposed procedure \eqref{eq:proposed}, define the constrained optimization problem
\begin{equation}
  \label{eq:proposed-i}
  \hat{t}_\rho^{(-i)}
  =
  \inf \left[ t \in [0, \infty) : \frac{\prod_{d = 1}^2 n^{-1} \{\sum_{j \ne i} I(T_{jd} \geq t) + 1\} + \rho}{n^{-1} \vee n^{-1} \{\sum_{j \ne i} I(T_{j1} \geq t,  T_{j2} \geq t) + 1\}} \leq \alpha \right].
\end{equation}
This type of leave-one-out construction of $\hat{t}_\rho^{(-i)}$ has also been used in proofs of false discovery rate control in a single sequence of test statistics \citep{sarkar2008methods, ramdas2017unified, benditkis2018false}. 

For any feature $i \in \mathcal{R}(\hat{t}_\rho)$, $I(T_{id} \geq \hat{t}_\rho) = 1$ and $I(T_{i1} \geq \hat{t}_\rho, T_{i2} \geq \hat{t}_\rho) = 1$, so
\[
\frac{\hat{S}_1(\hat{t}_\rho) \hat{S}_2(\hat{t}_\rho) + \rho}{n^{-1} \vee n^{-1} \sum_{i = 1}^n I(T_{i1} \geq \hat{t}_\rho, T_{i2} \geq \hat{t}_\rho)}
=
\frac{\prod_{d = 1}^2 n^{-1} \{\sum_{j \ne i} I(T_{jd} \geq \hat{t}_\rho) + 1\} + \rho}{n^{-1} \vee n^{-1} \{\sum_{j \ne i} I(T_{j1} \geq \hat{t}_\rho, T_{j2} \geq \hat{t}_\rho) + 1\}}.
\]
This means that $\hat{t}_\rho$ is feasible for problem \eqref{eq:proposed-i}, so $\hat{t}_\rho^{(-i)} \leq \hat{t}_\rho$. Next, this in turn implies that since $i \in \mathcal{R}(\hat{t}_\rho)$,  $1 = I(T_{id} \geq \hat{t}_\rho) \leq I(T_{id} \geq \hat{t}_\rho^{-(i)}) \leq 1$ and $1 = I(T_{i1} \geq \hat{t}_\rho, T_{i2} \geq \hat{t}_\rho) \leq I(T_{i1} \geq \hat{t}_\rho^{-(i)}, T_{i2} \geq \hat{t}_\rho^{-(i)}) \leq 1$, hence,  $I(T_{id} \geq \hat{t}_\rho) = I(T_{id} \geq \hat{t}_\rho^{-(i)}) = 1$ and $I(T_{i1} \geq \hat{t}_\rho, T_{i2} \geq \hat{t}_\rho) = I(T_{i1} \geq \hat{t}_\rho^{-(i)}, T_{i2} \geq \hat{t}_\rho^{-(i)}) = 1$, so
\[
\frac{\hat{S}_1(\hat{t}_\rho^{(-i)}) \hat{S}_2(\hat{t}_\rho^{(-i)}) + \rho}{n^{-1} \vee n^{-1} \sum_{i = 1}^n I(T_{i1} \geq \hat{t}_\rho^{(-i)}, T_{i2} \geq \hat{t}_\rho^{(-i)})}
=
\frac{\prod_{d = 1}^2 n^{-1} \{\sum_{j \ne i} I(T_{jd} \geq \hat{t}_\rho^{(-i)}) + 1\} + \rho}{n^{-1} \vee n^{-1} \{\sum_{j \ne i} I(T_{j1} \geq \hat{t}_\rho^{(-i)}, T_{j2} \geq \hat{t}_\rho^{(-i)}) + 1\}},
\]
which is at most $\alpha$ by construction of $\hat{t}_\rho^{(-i)}$. Thus $\hat{t}_\rho^{(-i)}$ is feasible for problem \eqref{eq:proposed} and $\hat{t}_\rho \leq \hat{t}_\rho^{(-i)}$.

The previous results imply that $\hat{t}_\rho^{(-i)} = \hat{t}_\rho$ for $i \in \mathcal{R}(\hat{t}_\rho)$. Then
\begin{align*}
  \textsc{fdr}(\hat{\delta}_\rho)
  \leq\,&
  \frac{\alpha}{n}
  \sum_{i: \bI_i \in \mathcal{S}_2^c \cap \mathcal{R}(\hat{t}_\rho)}
  E \frac{I(T_{i1} \geq \hat{t}_\rho, T_{i2} \geq \hat{t}_\rho)}{\hat{S}_1(\hat{t}_\rho) \hat{S}_2(\hat{t}_\rho) + \rho} \\
  \leq\,&
  \frac{\alpha}{n}
  \sum_{i: \bI_i \in \mathcal{S}_2^c \cap \mathcal{R}(\hat{t}_\rho)}
  E \frac{ I(T_{i1} \geq \hat{t}_\rho^{(-i)}, T_{i2} \geq \hat{t}_\rho^{(-i)})}{\hat{S}_1(\hat{t}_\rho^{(-i)}) \hat{S}_2(\hat{t}_\rho^{(-i)}) + \rho}
  \\
  \leq\,&
  \frac{\alpha}{n}
  \sum_{i: \bI_i \in \mathcal{S}_2^c \cap \mathcal{R}(\hat{t}_\rho)}
  E \frac{I(T_{i1} \geq \hat{t}_\rho^{(-i)}, T_{i2} \geq \hat{t}_\rho^{(-i)})}{\prod_{d = 1}^2 \{n^{-1} \sum_{j \ne i} I(T_{jd} \geq \hat{t}_\rho^{(-i)}) + 1\} + \rho},
\end{align*}
where the third line follows because it was shown above that $i \in \mathcal{R}(\hat{t}_\rho)$ implies $i \in \mathcal{R}(\hat{t}_\rho^{(-i)})$. Since neither $\hat{t}_\rho^{(-i)}$ nor the denominator of the final expression depends on $(T_{i1}, T_{i2})$, and because the $T_{id}$ are independent across sequences $d$, for every $i \in \mathcal{R}(\hat{t}_\rho)$
\begin{align*}
  E \frac{I(T_{i1} \geq \hat{t}_\rho^{(-i)}, T_{i2} \geq \hat{t}_\rho^{(-i)})}{\prod_{d = 1}^2 \{n^{-1} \sum_{j \ne i} I(T_{jd} \geq \hat{t}_\rho^{(-i)}) + 1\} + \rho}
  =\,&
  E \frac{S_{i1}^{I_{i1}}(\hat{t}_\rho^{(-i)}) S_{i2}^{I_{i2}}(\hat{t}_\rho^{(-i)})}{\prod_{d = 1}^2 \{n^{-1} \sum_{j \ne i} I(T_{jd} \geq \hat{t}_\rho^{(-i)}) + 1\} + \rho} \\
  =\,&
  E \frac{S_{i1}^{I_{i1}}(\hat{t}_\rho^{(-i)}) S_{i2}^{I_{i2}}(\hat{t}_\rho^{(-i)})}{\hat{S}_1(\hat{t}_\rho^{(-i)}) \hat{S}_2(\hat{t}_\rho^{(-i)}) + \rho}.
\end{align*}
Then again because $\hat{t}_\rho = \hat{t}_\rho^{(-i)}$ on $\mathcal{R}(\hat{t}_\rho)$,
\[
\textsc{fdr}(\hat{\delta}_\rho)
\leq
\frac{\alpha}{n}
\sum_{i: \bI_i \in \mathcal{S}_2^c \cap \mathcal{R}(\hat{t}_\rho)}
E \frac{S_{i1}^{I_{i1}}(\hat{t}_\rho) S_{i2}^{I_{i2}}(\hat{t}_\rho)}{\hat{S}_1(\hat{t}_\rho) \hat{S}_2(\hat{t}_\rho) + \rho}
\leq
\alpha
E \frac{n^{-1} \sum_{i: \bI_i \in \mathcal{S}_2^c} S_{i1}^{I_{i1}}(\hat{t}_\rho) S_{i2}^{I_{i2}}(\hat{t}_\rho)}{\hat{S}_1(\hat{t}_\rho) \hat{S}_2(\hat{t}_\rho) + \rho}.
\]

It remains to show that
\[
\limsup_{n \rightarrow \infty}
E \frac{n^{-1} \sum_{i: \bI_i \in \mathcal{S}_2^c} S_{i1}^{I_{i1}}(\hat{t}_\rho) S_{i2}^{I_{i2}}(\hat{t}_\rho)}{\hat{S}_1(\hat{t}_\rho) \hat{S}_2(\hat{t}_\rho) + \rho}
\leq 1.
\]
By the Fatou-Lebesgue theorem, it suffices to show that
\[
\limsup_{n \rightarrow \infty}\frac{n^{-1} \sum_{i: \bI_i \in \mathcal{S}_2^c} S_{i1}^{I_{i1}}(\hat{t}_\rho) S_{i2}^{I_{i2}}(\hat{t}_\rho)}{\hat{S}_1(\hat{t}_\rho) \hat{S}_2(\hat{t}_\rho) + \rho}
\leq
1
\]
almost surely. The left-hand expression can be rewritten as
\begin{align}
  &
  \limsup_{n \rightarrow \infty}
  \frac{n^{-1} \sum_{i: \bI_i \in \mathcal{S}_2^c} S_{i1}^{I_{i1}}(\hat{t}_\rho) S_{i2}^{I_{i2}}(\hat{t}_\rho) }{\hat{S}_1(\hat{t}_\rho) \hat{S}_2(\hat{t}_\rho) + \rho}
  \nonumber \\
  \leq\,&
  1
  +
  \limsup_{n \rightarrow \infty}
  \frac{n^{-1} \sum_{i: \bI_i \in \mathcal{S}_2^c} S_{i1}^{I_{i1}}(\hat{t}_\rho) S_{i2}^{I_{i2}}(\hat{t}_\rho) - \sum_{\bI \in \mathcal{S}_2^c} \pi_{\bI} S_1^{I_1}(\hat{t}_\rho) S_2^{I_2}(\hat{t}_\rho)}{\hat{S}_1(\hat{t}_\rho) \hat{S}_2(\hat{t}_\rho) + \rho}
  \,+
  \label{eq:1} \\
  &
  \limsup_{n \rightarrow \infty}
  \frac{\sum_{\bI \in \mathcal{S}_2^c} \pi_{\bI} S_1^{I_1}(\hat{t}_\rho) S_2^{I_2}(\hat{t}_\rho) - S_1(\hat{t}_\rho) S_2(\hat{t}_\rho) - \rho}{\hat{S}_1(\hat{t}_\rho) \hat{S}_2(\hat{t}_\rho) + \rho}
  \,+
  \label{eq:2} \\
  &
  \limsup_{n \rightarrow \infty}
  \frac{S_1(\hat{t}_\rho) S_2(\hat{t}_\rho) - \hat{S}_1(\hat{t}_\rho) \hat{S}_2(\hat{t}_\rho)}{\hat{S}_1(\hat{t}_\rho) \hat{S}_2(\hat{t}_\rho) + \rho},
  \label{eq:3}
\end{align}
with $\pi_{\bI}$ and $S_d(\hat{t}_d)$ defined as in Assumption \ref{a:conv}.

First, the second term of \eqref{eq:1} obeys
\begin{align*}
  &
  \limsup_{n \rightarrow \infty}
  \left \vert
  \frac{n^{-1} \sum_{i: \bI_i \in \mathcal{S}_2^c} S_{i1}^{I_{i1}}(\hat{t}_\rho) S_{i2}^{I_{i2}}(\hat{t}_\rho) - \sum_{\bI \in \mathcal{S}_2^c} \pi_{\bI} S_1^{I_1}(\hat{t}_\rho) S_2^{I_2}(\hat{t}_\rho)}{\hat{S}_d(\hat{t}_\rho) \hat{S}_d(\hat{t}_\rho) + \rho}
  \right \vert
  \\
  \leq\,&
  \frac{1}{\rho}
  \lim_{n \rightarrow \infty} \sup_{t \in [0, \infty)}
  \left\vert
  n^{-1} \sum_{i: \bI_i \in \mathcal{S}_2^c} S_{i1}^{I_{i1}}(t) S_{i2}^{I_{i2}}(t) - \sum_{\bI \in \mathcal{S}_2^c} \pi_{\bI} S_1^{I_1}(t) S_2^{I_2}(t)
  \right\vert
  =
  0,
\end{align*}
almost surely, by Assumption \ref{a:conv}. Next, the numerator of \eqref{eq:2} satisfies
\[
\sup_{t \in [0, \infty)} \left\{ \sum_{\bI \in \mathcal{S}_2^c} \pi_{\bI} S_1^{I_1}(t) S_2^{I_2}(t) - S_1(t) S_2(t) - \rho \right\}
<
0
\]
by Proposition \ref{prop:bound}, and because $\rho > 0$,
\[
\limsup_{n \rightarrow \infty}
\frac{\sum_{\bI \in \mathcal{S}_2^c} \pi_{\bI} S_1^{I_1}(\hat{t}_\rho) S_2^{I_2}(\hat{t}_\rho) - S_1(\hat{t}_\rho) S_2(\hat{t}_\rho) - \rho}{\hat{S}_1(\hat{t}_\rho) \hat{S}_2(\hat{t}_\rho) + \rho}
\leq
0
\]
almost surely. It remains to show that \eqref{eq:3} goes to zero. Since
\begin{align*}
  \limsup_{n \rightarrow \infty}
  \left \vert
  \frac{S_1(\hat{t}_\rho) S_2(\hat{t}_\rho) - \hat{S}_1(\hat{t}_\rho) \hat{S}_2(\hat{t}_\rho)}{\hat{S}_1(\hat{t}_\rho) \hat{S}_2(\hat{t}_\rho) + \rho}
  \right \vert
  \leq
  \frac{1}{\rho}
  \lim_{n \rightarrow \infty} \sup_{t \in [0, \infty)}
  \left \vert S_1(t) S_2(t) - \hat{S}_1(t) \hat{S}_2(t) \right\vert
\end{align*}
and the $\hat{S}_d(t)$ are averages of independent but non-identically distributed terms that satisfy the conditions of Theorem 8.3 of \citet{pollard1990empirical},
\[
\lim_{n \rightarrow \infty}
\sup_{t \in [0, \infty)} \left \vert \hat{S}_d(t) - \frac{1}{n} \sum_{i = 1}^n S_{id}^{I_{id}}(t) \right \vert
=
0
\]
almost surely for all $d$. By Assumption \ref{a:conv},
\[
\lim_{n \rightarrow \infty}
\sup_{t \in [0, \infty)}
\left\vert
\frac{1}{n} \sum_{i = 1}^n S_{id}^{I_{id}}(t)
-
S_d(t)
\right\vert
=
0,
\]
where 
$S_d(t) = (1 - \pi_d) S_d^0(t) + \pi_d S_d^1(t)$ is the marginal survival function defined in Proposition \ref{prop:bound}. Therefore 
\begin{align*}
  \lim_{n \rightarrow \infty} \sup_{t \in [0, \infty)}
  \vert S_1(t) S_2(t) - \hat{S}_1(t) \hat{S}_2(t) \vert
  \leq\,&
  \lim_{n \rightarrow \infty} \sup_{t \in [0, \infty)} \vert S_1(t) S_2(t) - \hat{S}_1(t) S_2(t) \vert +\,\\
  &
  \lim_{n \rightarrow \infty} \sup_{t \in [0, \infty)} \vert \hat{S}_1(t) S_2(t) - \hat{S}_1(t) \hat{S}_2(t) \vert \\
  \leq\,&
  \lim_{n \rightarrow \infty} \sum_{d = 1}^2 \sup_{t \in [0, \infty)} \vert S_d(t) - \hat{S}_d(t) \vert
  = 0
\end{align*}
almost surely. This concludes the proof.

\section{Proof of Theorem \ref{thm:alt_conservative}}
Define
\begin{equation}
  \label{eq:fxns}
  \begin{aligned}
    V_{ab}(t_1, t_2)
    &=
    \sum_{i:I_{i1}=a, I_{i2}=b} I(T_{i2} \geq t_1, T_{i2} \geq t_2),
    \quad
    a,b = 0,1,
    \\
    R(t_1, t_2)
    &=
    \sum_i I(T_{i1} \geq t_1, T_{i2} \geq t_2).
  \end{aligned}
\end{equation}
Then the true false discovery rate attained by a discovery rule of the form $\delta(T_{i1}, T_{i2}) = I(T_{i1} \geq t_1, T_{i2} \geq t_2)$ can be written as
\[
\textsc{fdr}(\delta)
=
E\left[ \frac{V_{00}(t_1,t_2)+V_{10}(t_1,t_2)+V_{01}(t_1,t_2)}{\max\{1,R(t_1,t_2)\}} \right].
\]

It will first be shown that for any $\eta_1,\eta_2 < \infty$,
\begin{equation}
  \label{eq:1.1}
  \lim_{n \rightarrow \infty}
  \inf_{t_1 \leq \eta_1, t_2 \leq \eta_2}
  \left[
  \widetilde{\textsc{fdp}}_\rho(t_1,t_2) - \frac{n^{-1}}{n^{-1}}\frac{V_{00}(t_1,t_2)+V_{10}(t_1,t_2)+V_{01}(t_1,t_2)}{\max\{1, R(t_1,t_2)\}}
  \right]
  \geq0
\end{equation}
almost surely, for $\widetilde{\textsc{fdp}}_\rho(t_1,t_2)$ defined in \eqref{eq:alt_fdp}. Next it will be shown that
\begin{equation}
  \label{eq:1.2}
  \sup_{t_1 \leq \eta_1, t_2 \leq \eta_2}
  \left\vert\frac{n^{-1}}{n^{-1}}\frac{V_{00}(t_1,t_2)+V_{10}(t_1,t_2)+V_{01}(t_1,t_2)}{\max\{1, R(t_1,t_2)\}}-\textsc{fdr}(\delta)\right\vert
  \rightarrow
  0,
\end{equation}
almost surely, which will complete the proof.

To show~\eqref{eq:1.1}, it suffices to show
\[
\lim_{n \rightarrow \infty}
\inf_{t_1 \leq\eta_1, t_2 \leq \eta_2}
[
\hat{S}_1(t_1)\hat{S}_2(t_2)
-
n^{-1}\{V_{00}(t_1,t_2)+V_{10}(t_1,t_2)+V_{01}(t_1,t_2)\}
]
\geq
0
\]
almost surely. Arguments from the proof of Theorem \ref{thm:fdr} can be used to show that
\begin{align*}
  &
  \lim_{n \rightarrow \infty} \sup_{t_1,t_2\in [0, \infty)}
  \vert\hat{S}_1(t_1)\hat{S}_2(t_2)-S_1(t_1)S_2(t_2)\vert = 0, \\
  &
  \lim_{n \rightarrow \infty} \sup_{t_1,t_2\in [0, \infty)}
  \vert
  G^0(t_1, t_2)
  -
  n^{-1} \{V_{00}(t_1,t_2) + V_{10}(t_1,t_2) + V_{01}(t_1,t_2)\}
  \vert
  =
  0
\end{align*}
almost surely, where $G^0(t_1, t_2) = \sum_{\bI \in \mathcal{S}_2^c} \pi_{\bI} S_1^{I_1}(t_1) S_2^{I_2}(t_2)$. Combining these with Proposition \ref{prop:bound} proves~\eqref{eq:1.1}.

To prove \eqref{eq:1.2}, define
\begin{equation}
  \label{eq:G}
  G(t_1, t_2) = G^0(t_1, t_2) + \pi_{(1,1)} G^1(t_1,t_2)
\end{equation}
for $G^1(t_1, t_2)$ from in Assumption \ref{a:conv}. Then
\begin{align*}
  &
  \lim_{n\rightarrow\infty} \sup_{t_1 \leq \eta_1, t_2 \leq \eta_2}
  \left\vert
  \frac{n^{-1}}{n^{-1}}
  \frac{V_{00}(t_1,t_2)+V_{10}(t_1,t_2)+V_{01}(t_1,t_2)}
       {\max\{1, R(t_1,t_2)\}}
  -
  \frac{G^0(t_1,t_2)}
       {G(t_1,t_2)}
  \right\vert\\
  \leq\,&
  \lim_{n\rightarrow\infty}
  \frac{n}{\max\{1, R(\eta_1,\eta_2)\}}
  \sup_{t_1 \leq \eta_1, t_2 \leq \eta_2}
  \left\vert
  n^{-1}\{V_{00}(t_1,t_2)+V_{10}(t_1,t_2)+V_{01}(t_1,t_2)\}
  -
  G^0(t_1,t_2)
  \right\vert
  \,+\\
  &
  \lim_{n\rightarrow\infty}
  \frac{n}{\max\{1, R(\eta_1,\eta_2)\}}\frac{1}{G(\eta_1,\eta_2)}
  \sup_{t_1 \leq \eta_1, t_2 \leq \eta_2}
  \left\vert
  G(t_1,t_2)-n^{-1}\max\{1, R(t_1,t_2)\}
  \right\vert.
\end{align*}
Arguments from the proof of Theorem \ref{thm:fdr} can be used to show that both terms on the right-hand side equal zero almost surely. Next, the dominated convergence theorem implies that
\begin{align*}
  0
  =\,&
  E\lim_{n \rightarrow \infty} \sup_{t_1 \leq \eta_1, t_2 \leq \eta_2}
  \left\vert
  \frac{n^{-1}}{n^{-1}}
  \frac{V_{00}(t_1,t_2)+V_{10}(t_1,t_2)+V_{01}(t_1,t_2)}
       {\max\{1, R(t_1,t_2)\}}
  -
  \frac{G^0(t_1,t_2)}
       {G(t_1,t_2)}
  \right\vert\\
  =\,&
  \lim_{n \rightarrow \infty}
  E \sup_{t_1 \leq \eta_1, t_2 \leq \eta_2}
  \left\vert
  \frac{n^{-1}}{n^{-1}}
  \frac{V_{00}(t_1,t_2)+V_{10}(t_1,t_2)+V_{01}(t_1,t_2)}
       {\max\{1, R(t_1,t_2)\}}
  -
  \frac{G^0(t_1,t_2)}
       {G(t_1,t_2)}
  \right\vert\\
  \geq\,&
  \lim_{n \rightarrow \infty} \sup_{t_1 \leq \eta_1, t_2 \leq \eta_2}
  \left\vert
  \textsc{fdr}(\delta)
  -
  \frac{G^0(t_1,t_2)}
       {G(t_1,t_2)}
  \right\vert.
\end{align*}
Combining these results proves \eqref{eq:1.2}.

\section{Proof of Theorem \ref{thm:alt_fdr}}
The theorem is trivially true when $(\hat{t}_{\rho 1},\hat{t}_{\rho 2}) = (\infty, \infty)$. Otherwise, suppose there exist fixed $\eta_1, \eta_2 < \infty$ such that $\hat{t}_{\rho 1} \leq \eta_1$ and $\hat{t}_{\rho 2} \leq \eta_2$ with probability 1. Then by~\eqref{eq:1.1} from the proof of Theorem~\ref{thm:alt_conservative},
\begin{align*}
  &
  \liminf_{n\rightarrow\infty}
  \left[
  \widetilde{\textsc{fdp}}_\rho(\hat{t}_{\rho 1},\hat{t}_{\rho 2})
  -
  \frac{
    V_{00}(\hat{t}_{\rho 1},\hat{t}_{\rho 2})+V_{10}(\hat{t}_{\rho 1},\hat{t}_{\rho 2})+V_{01}(\hat{t}_{\rho 1},\hat{t}_{\rho 2})
  }{
    \max\{1, R(\hat{t}_{\rho 1},\hat{t}_{\rho 2})\}
  }
  \right ]
  \\
  \geq\,&
  \lim_{n \rightarrow \infty}
  \inf_{t_1 \leq \eta_1, t_2 \leq \eta_2}
  \left[
  \widetilde{\textsc{fdp}}_\rho(t_1,t_2) - \frac{n^{-1}}{n^{-1}}\frac{V_{00}(t_1,t_2)+V_{10}(t_1,t_2)+V_{01}(t_1,t_2)}{\max\{1, R(t_1,t_2)\}}
  \right]
  \geq0
\end{align*}
almost surely. This implies that
\[
\limsup_{n\rightarrow\infty}
\frac{V_{00}(\hat{t}_{\rho 1},\hat{t}_{\rho 2})+V_{10}(\hat{t}_{\rho 1},\hat{t}_{\rho 2})+V_{01}(\hat{t}_{\rho 1},\hat{t}_{\rho 2})}
     {\max\{1, R(\hat{t}_{\rho 1},\hat{t}_{\rho 2})\}}
\leq
\alpha
\]
almost surely. Then by the Fatou-Lebesgue theorem,
\[
\limsup_{p\rightarrow\infty}
\textsc{fdr}(\tilde{\delta})
\leq
E
\limsup_{p\rightarrow\infty}
\frac{V_{00}(\hat{t}_{\rho 1},\hat{t}_{\rho 2})+V_{10}(\hat{t}_{\rho 1},\hat{t}_{\rho 2})+V_{01}(\hat{t}_{\rho 1},\hat{t}_{\rho 2})}
     {\max\{1, R(\hat{t}_{\rho 1},\hat{t}_{\rho 2})\}}
\leq
\alpha
\]
for the discovery procedure $\tilde{\delta}$ \eqref{eq:alt}.

It remains to construct $\eta_1$ and $\eta_2$. The pointwise limit of $\widetilde{\textsc{fdp}}_\rho(t_1, t_2)$ \eqref{eq:alt_fdp} is
\[
\textsc{fdp}_\rho(t_1,t_2) = \{S_1(t_1) S_2(t_2) + \rho\} / G(t_1,t_2),
\]
for $G(t_1, t_2)$ defined in \eqref{eq:G}. By assumption, there exists some $\epsilon > 0$ such that $\textsc{fdp}_\rho(t'_1, t'_2) = \alpha - \epsilon$. Kolmogorov's strong law of large numbers and Slutsky's theorem show that for $n$ sufficiently large,
\[
\vert\widetilde{\textsc{fdp}}_\rho(t'_1,t'_2)-\textsc{fdp}_\rho(t'_1,t'_2)\vert\leq\epsilon/2
\]
with probability 1. This implies that $\widetilde{\textsc{fdp}}_\rho(t'_1,t'_2)\leq\alpha-\epsilon/2$, so $(t'_1,t'_2)$ is a feasible solution of the optimization problem \eqref{eq:alt}. Then $\hat{G}(\hat{t}_{\rho 1},\hat{t}_{\rho 2})\geq\hat{G}(t'_1,t'_2)$. Using arguments from the proof of Theorem \ref{thm:fdr}, it can be shown that $\hat{G}(t_1, t_2)$ converges almost surely to $G(t_1, t_2)$ uniformly in $(t_1, t_2)$. Therefore for any $\eta > 0$, there exists a sufficiently large $n$ such that
\[
G(\hat{t}_{\rho 1}, \hat{t}_{\rho 2})
\geq
\hat{G}(\hat{t}_{\rho 1}, \hat{t}_{\rho 2}) - \eta / 4
\geq
\hat{G}(t'_1, t'_2) - \eta /4
\geq
G(t'_1, t'_2) - \eta / 2
\]
with probability 1. Choose $\eta = G(t'_1, t'_2)$, which must be positive because $t'_1$ and $t'_2$ are both finite by assumption. This shows that $G(\hat{t}_{\rho 1}, \hat{t}_{\rho 2}) \geq \eta / 2 > 0$ with probability 1. Now define $\eta_1$ such that $S_1^{-1}(\eta / 2)$ and $\eta_2 = S_2^{-1}(\eta / 2)$. Then
\[
G(\hat{t}_{\rho 1}, \hat{t}_{\rho 2})
\geq
\eta / 2
=
S_1(\eta_1)
=
G(\eta_1, 1)
\geq
G(\eta_1, \hat{t}_{\rho 2}),
\]
which implies that $\hat{t}_{\rho 1} \leq \eta_1$ with probability 1. By similar reasoning, $\hat{t}_{\rho 2} \leq \eta_2$ with probability 1. Finally, since $\eta > 0$, $\eta_1$ and $\eta_2$ are both finite as well.

\section{Proof of Proposition \ref{prop:bound>=2}}
We prove  the Proposition  \ref{prop:bound>=2} for $D = 3$. Similar arguments can be applied for cases $D \geq 4$.
First, the expression
\[
\sum_{\bI \in \mathcal{S}_3^c} \pi_{\bI} S_1^{I_1}(t_1) S_2^{I_2}(t_2) S_3^{I_3}(t_3)
\]
equals
\begin{align*}
  &
  \pi_{(0, 0, 0)} S_1^0(t_1) S_2^0(t_2) S_3^0(t_3) +
  \pi_{(0, 1, 0)} S_1^0(t_1) S_2^1(t_2) S_3^0(t_3) +
  \pi_{(1, 0, 1)} S_1^1(t_1) S_2^0(t_2) S_3^1(t_3) \,+\\
  &
  \pi_{(1, 0, 0)} S_1^1(t_1) S_2^0(t_2) S_3^0(t_3) +
  \pi_{(0, 0, 1)} S_1^0(t_1) S_2^0(t_2) S_3^1(t_3) +
  \pi_{(0, 1, 0)} S_1^0(t_1) S_2^1(t_2) S_3^0(t_3)\,+\\
  &
  \pi_{(1, 1, 0)} S_1^1(t_1) S_2^1(t_2) S_3^0(t_3).
\end{align*}
By the stochastic ordering in Assumption \ref{a:sto_ord}, $S_d^0(t_d) < S_d^1(t_d)$ for $d = 1, 2, 3$, so the previous expression can be upper-bounded by
\begin{align*}
  &
  \{\pi_{(0, 0, 0)} S_1^0(t_1) S_2^0(t_2) +
  \pi_{(0, 1, 0)} S_1^0(t_1) S_2^1(t_2)  +
  \pi_{(1, 0, 1)} S_1^1(t_1) S_2^0(t_2)\} S_3^1(t_3) \,+\\
  &
  \{\pi_{(1, 0, 0)} S_2^0(t_2) S_3^0(t_3) +
  \pi_{(0, 0, 1)} S_2^0(t_2) S_3^1(t_3) +
  \pi_{(0, 1, 0)}  S_2^1(t_2) S_3^0(t_3)\} S_1^1(t_1) \,+\\
  &
  \pi_{(1, 1, 0)} S_1^1(t_1) S_2^1(t_2) S_3^0(t_3) \\
  \leq\,&
  \{\pi_{(0, 0, 0)} S_1^0(t_1) S_2^0(t_2) +
  \pi_{(0, 1, 0)} S_1^0(t_1) S_2^1(t_2)  +
  \pi_{(1, 0, 1)} S_1^1(t_1) S_2^0(t_2)\} \,+\\
  &
  \{\pi_{(1, 0, 0)} S_2^0(t_2) S_3^0(t_3) + 
  \pi_{(0, 0, 1)} S_2^0(t_2) S_3^1(t_3) +
  \pi_{(0, 1, 0)}  S_2^1(t_2) S_3^0(t_3)\} \,+\\
  &
  \{\pi_{(0, 1, 0)} S_1^0(t_1)S_3^0(t_3) + 
  \pi_{(0, 1, 1)} S_1^0(t_1)S_3^1(t_3) +
  \pi_{(1, 1, 0)} S_1^1(t_1)S_3^0(t_3)\}.
\end{align*}
Now define $\pi_{(\theta_1, \theta_2, \cdot)} = \pi_{(\theta_1, \theta_2, 0)} + \pi_{(\theta_1, \theta_2, 1)}$ for $\theta_1, \theta_2 \in \{0, 1\}$, and define $\pi_{(\theta_1, \cdot, \theta_2)}$ and $\pi_{(\cdot, \theta_1, \theta_2)}$ similarly. Then the previous expression is upper-bounded by
\begin{align*}
  &
  \{\pi_{(0, 0, \cdot)} S_1^0(t_1) S_2^0(t_2) +
  \pi_{(0, 1,  \cdot)} S_1^0(t_1) S_2^1(t_2) +
  \pi_{(1, 0, \cdot)} S_1^1(t_1) S_2^0(t_2)\} \,+\\
  &
   \{\pi_{(\cdot, 0, 0)} S_2^0(t_2) S_3^0(t_3) +
  \pi_{(\cdot, 0, 1)} S_2^0(t_2) S_3^1(t_3) +
  \pi_{(\cdot, 1, 0)} S_2^1(t_2) S_3^0(t_3) \}\,+\\
  &
  \{\pi_{(0, \cdot, 0)} S_1^0(t_1) S_3^0(t_3) +
  \pi_{(0, \cdot, 1)} S_1^0(t_1) S_3^1(t_3) +
  \pi_{(1, \cdot, 0)} S_1^1(t_1) S_3^0(t_3)\}.
\end{align*}
Applying Proposition \ref{prop:bound} to each of these terms gives the desired result.

\end{document}